\documentclass[twocolumn,aps,prl] {revtex4} 
\usepackage{graphicx}
\begin{document}
\pagestyle{empty} 
\title{Leak-rate of seals: effective medium theory and comparison with experiment}
\author{  
B. Lorenz and B.N.J. Persson} 
\affiliation{IFF, FZ J\"ulich, D-52425 J\"ulich, Germany}

\begin{abstract}
Seals are extremely useful devices to prevent fluid leakage. 
We present an effective medium theory of the leak-rate of rubber seals,
which is based on a recently developed contact mechanics theory. 
We compare the theory with experimental results for seals consisting of silicon rubber 
in contact with sandpaper and sand-blasted PMMA surfaces.
\end{abstract}
\maketitle


{\bf 1. Introduction}

A seal is a device for closing a gap or making a joint fluid tight\cite{Flitney}.
Seals play a crucial role in many modern engineering devices, and the failure of
seals may result in catastrophic events, such as the Challenger disaster. 
In spite of its apparent 
simplicity, it is not easy to predict the leak-rate and
(for dynamic seals) the friction forces\cite{Mofidi}. 
The main problem is the influence of surface
roughness on the contact mechanics at the seal-substrate interface. 
Most surfaces of engineering interest have surface roughness
on a wide range of length scales\cite{P3}, e.g, from cm to nm, which will influence the leak rate
and friction of seals, and accounting for the whole range of surface roughness
is impossible using standard numerical methods, such as the Finite Element Method.

\begin{figure}
\includegraphics[width=0.45\textwidth,angle=0]{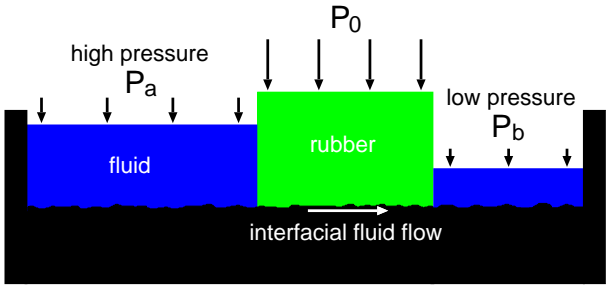}
\caption{\label{water.rubber}
Rubber seal (schematic). The liquid on the left-hand-side is under the hydrostatic
pressure $P_{\rm a}$ and the liquid to the right under the pressure $P_{\rm b}$
(usually, $P_{\rm b}$ is the atmospheric pressure).
The pressure difference $\Delta P = P_{\rm a}-P_{\rm b}$ results in liquid flow
at the interface between the rubber seal and the rough substrate surface. The
volume of liquid flow per unit time is denoted by $\dot Q$, and depends on the 
squeezing pressure $P_0$ acting on the rubber seal. 
}
\end{figure}

We have recently presented experimental results for the leak-rate of rubber seals\cite{LorenzEPL},
and compared the results to a ``single-junction'' theory\cite{Creton,P3,Yang}, 
which is based on percolation theory and a recently developed contact mechanics 
theory\cite{JCPpers,PerssonPRL,PSSR,P1,Bucher,YangPersson,PerssonJPCM,earlier}.
Here we will report on new experimental data, and compare the
experimental results with the single-junction
theory, and also to an extension of this theory presented below, 
which is based on the effective medium approach.

\begin{figure}
\includegraphics[width=0.45\textwidth,angle=0.0]{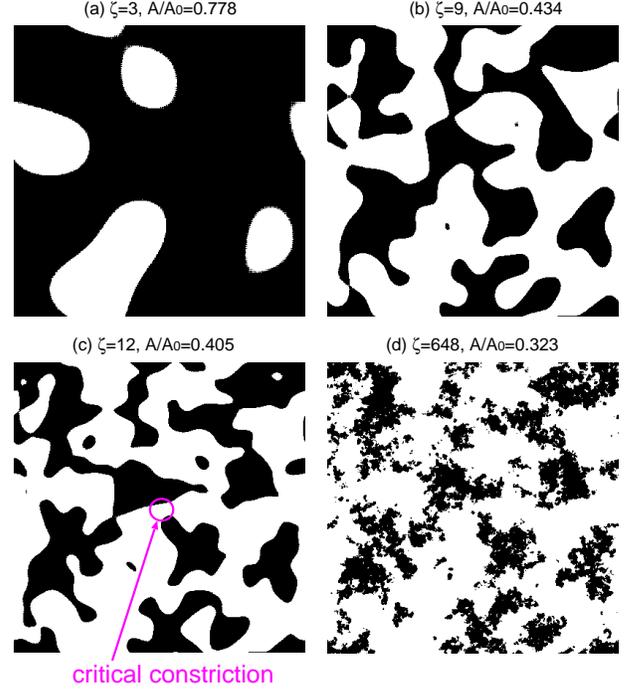}
\caption{\label{percolationpic}
The contact region at different magnifications $\zeta = 3$, 9, 12 and 648, is shown in
(a)-(d) respectively. 
When the magnification increases from 9 to 12 the non-contact region percolate.
At the lowest magnification $\zeta = 1$: $A(1)=A_0$. The figure is the result
of Molecular Dynamics simulations of the contact between elastic solids with randomly rough surfaces,
see Ref. \cite{Yang}.
}
\end{figure}

\vskip 0.3cm

{\bf 2. Single-junction theory}

We first briefly review the leak-rate model developed in Ref. \cite{Creton,P3,Yang}.
Consider the fluid leakage through a rubber seal, from a high fluid pressure $P_{\rm a}$ region, to a
low fluid pressure $P_{\rm b}$ region, as in Fig. \ref{water.rubber}. 
Assume that the nominal contact region between the rubber and the hard countersurface is
rectangular with area $L_x\times L_y$, with $L_y > L_x$. 
We assume that the high pressure fluid region is for $x<0$
and the low pressure region for $x>L_x$. We  ``divide'' the contact region into squares with
the side $L_x=L$ and the area $A_0=L^2$ (this assumes that $N=L_y/L_x$ is an integer, but this 
restriction does not affect the final result). 
Now, let us study the contact between the two solids within one of the squares
as we change the magnification $\zeta$. We define $\zeta= L/\lambda$, where $\lambda$ is the resolution.
We study how the apparent contact area (projected on the $xy$-plane),
$A(\zeta)$, between the two solids depends on the magnification $\zeta$.
At the lowest magnification we cannot observe any surface roughness, and 
the contact between the solids appears to be complete i.e., $A(1)=A_0$. 
As we increase the magnification
we will observe some interfacial roughness, and the (apparent) contact area will decrease.
At high enough magnification, say $\zeta = \zeta_{\rm c}$, a percolating path of 
non-contact area will be observed 
for the first time, see Fig.~\ref{percolationpic}. 
We denote the most narrow constriction along this percolation path as
the {\it critical constriction}. The critical constriction will have the lateral
size $\lambda_{\rm c} = L/\zeta_{\rm c}$ and the surface separation at this point is denoted by 
$u_{\rm c}$. We can calculate $u_{\rm c}$
using a recently developed contact mechanics theory\cite{YangPersson} (see below). 
As we continue to increase the magnification we will find more percolating channels 
between the surfaces, but these will have more narrow constrictions 
than the first channel which appears at $\zeta=\zeta_{\rm c}$, and as a first approximation one may
neglect the contribution to the leak-rate from these channels\cite{Yang}. 

A first rough estimate of the leak-rate is obtained by assuming that all the leakage 
occurs through the critical percolation channel, and that
the whole pressure drop $\Delta P = P_{\rm a}-P_{\rm b}$ (where $P_{\rm a}$ and $P_{\rm b}$ is the 
pressure to the left and right of the
seal) occurs over the critical constriction (of width and length $\lambda_{\rm c} \approx L/\zeta_{\rm c}$
and height $u_{\rm c}$). 
We will refer to this theory as the ``single-junction'' theory.
If we approximate the critical constriction
as a pore with rectangular cross section (width and length $\lambda_c$ and height $u_c << \lambda_c$),  
and if assume an incompressible
Newtonian fluid, the volume-flow per unit time through the critical constriction
will be given by (Poiseuille flow) 
$$\dot Q = {u_c^3(\zeta_{\rm c}) \over 12 \eta}  \Delta P,\eqno(1)$$
where $\eta $ is the fluid viscosity. 
In deriving (1) we have assumed laminar flow and that $u_c << \lambda_c$,
which is always satisfied in practice. We have also assumed no-slip boundary condition
on the solid walls. This assumption is not always satisfied at the micro or nano-scale, but is likely to be
a very good approximation in the present case owing to surface roughness which occurs at length-scales 
shorter than the size of the critical constriction. 
Finally, since there are
$N=L_y/L_x$ square areas in the rubber-countersurface (apparent) contact area, we get the total leak-rate
$$\dot Q = {L_y \over L_x} {u_c^3(\zeta_{\rm c}) \over 12 \eta}  \Delta P.\eqno(2)$$ 
Note that a given percolation channel could have several narrow (critical or nearly critical)
constrictions of nearly the same dimension
which would reduce the flow along the channel. But in this case one would also expect more channels from
the high to the low fluid pressure side of the junction, which would tend to increase the leak rate.
These two effects will, at least in the simplest picture where one assumes that the distance between the 
critical junctions along a percolation path (in the $x$-direction) is the same as the distance between the 
percolation channels (in the $y$-direction), compensate 
each other (see Ref. \cite{Yang}). 
The effective medium theory presented below
includes (in an approximate way) all the flow channels.

To complete the theory we must calculate the separation $u_{\rm c}$ 
of the surfaces at the
critical constriction. We first determine the critical magnification $\zeta_{\rm c}$ by assuming that the 
apparent relative contact area at this point is given by site percolation theory. 
Thus, the relative contact area $A(\zeta)/A_0 \approx 1-p_{\rm c}$, where $p_{\rm c}$  is the 
so called site percolation threshold\cite{Stauffer}. 
For an infinite-sized systems 
$p_{\rm c}\approx 0.696$ for a hexagonal lattice and $0.593$ for a square lattice\cite{Stauffer}. 
For finite sized systems the percolation will, on the average, occur for (slightly) smaller values
of $p$, and fluctuations in the percolation threshold will occur between 
different realizations of the same physical system. 
We take $p_{\rm c}\approx 0.6$ so that $A(\zeta_{\rm c})/A_0 \approx 0.4$ will determine the critical
magnification $\zeta=\zeta_{\rm c}$. 

The (apparent) relative contact area $A(\zeta)/A_0$ at the magnification $\zeta$
can be obtained using the contact mechanics 
formalism developed elsewhere\cite{PSSR,YangPersson,P1,Bucher,JCPpers},
where the system is studied at different magnifications $\zeta$.
We have\cite{JCPpers,PerssonPRL}

$${A(\zeta)\over A_0} = {1\over (\pi G )^{1/2}}\int_0^{P_0} d\sigma \ {\rm e}^{-\sigma^2/4G} 
= {\rm erf} \left ( P_0 \over 2 G^{1/2} \right )$$
where
$$G(\zeta) = {\pi \over 4}\left ({E\over 1-\nu^2}\right )^2 \int_{q_0}^{\zeta q_0} dq q^3 C(q)$$
where the surface roughness power spectrum
$$C(q) = {1\over (2\pi)^2} \int d^2x \langle h({\bf x})h({\bf 0})\rangle {\rm e}^{-i{\bf q}\cdot {\bf x}}$$
where $\langle ... \rangle$ stands for ensemble average. 
Here $E$ and $\nu$ are the Young's elastic modulus and the Poisson 
ratio of the rubber. The height profile $h({\bf x})$ of the rough surface can be measured routinely
today on all relevant length scales using optical and stylus experiments.

\begin{figure}
\includegraphics[width=0.45\textwidth,angle=0.0]{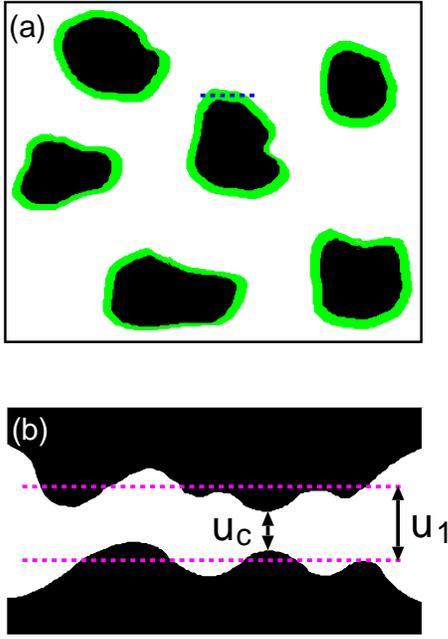}
\caption{\label{pic.Az.Azdz.rough}
(a) The black area is the asperity contact regions at the magnification $\zeta$.
The green area is the additional contact area observed when the magnification is
reduced to $\zeta-\Delta \zeta$ (where $\Delta \zeta$ is small). The average separation
between the solid walls in the green surface 
area is denoted by $u_1(\zeta)$. (b) The separation
between the solid walls along the blue dashed line in (a).
Since the surfaces of the solids are everywhere rough the actual 
separation between the solid walls in the green area
will fluctuate around the average $u_1(\zeta)$. At the most narrow constriction
the surface separation is $u_c$.
}
\end{figure}

\begin{figure}
\includegraphics[width=0.35\textwidth,angle=0.0]{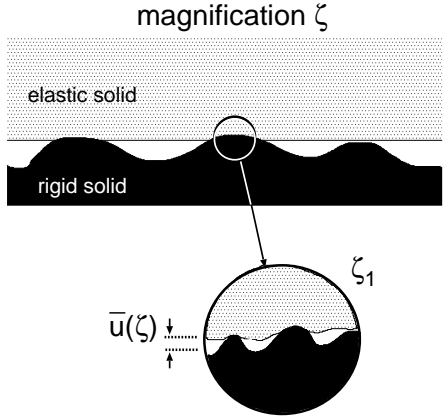}
\caption{\label{asperity.mag}
An asperity contact region observed at the magnification $\zeta$. It appears that
complete contact occur in the asperity contact region, but when the magnification is
increasing to the highest (atomic scale) magnification $\zeta_1$, 
it is observed that the solids are actually separated by the average distance $\bar{u}(\zeta)$.
}
\end{figure}

We define $u_1(\zeta)$ to be the (average) height separating the surfaces which appear to come into 
contact when the magnification decreases from $\zeta$ to $\zeta-\Delta \zeta$, where $\Delta \zeta$
is a small (infinitesimal) change in the magnification. 
In Fig. \ref{pic.Az.Azdz.rough}(a)
the black area is the asperity contact regions at the magnification $\zeta$.
The green area is the additional contact area observed when the magnification is
reduced to $\zeta-\Delta \zeta$ (where $\Delta \zeta$ is small)\cite{complex}. 
The average separation
between the solid walls in the green surface 
area is given by $u_1(\zeta)$. 
Fig. \ref{pic.Az.Azdz.rough}(b) shows the separation
between the solid walls along the dashed line in Fig. \ref{pic.Az.Azdz.rough}(a).
Since the surfaces of the solids are everywhere rough the actual 
separation between the solid walls in the green area
will fluctuate around the average $u_1(\zeta)$. Thus we expect $u_c=\alpha u_1(\zeta_c)$, where
$\alpha < 1$ (but of order unity, see Fig. \ref{pic.Az.Azdz.rough}(b))\cite{WithYang}. 
We note that $\alpha$ is due to the surface 
roughness which occur at length scales shorter
than $\lambda_c$, and it may be possible to calculate (or estimate) 
$\alpha$ from the surface roughness power spectrum,
but no such theory has been developed so far and here we treat $\alpha$ as a fit parameter.

$u_1(\zeta)$ is a monotonically decreasing
function of $\zeta$, and can be calculated from the average interfacial separation
$\bar u(\zeta)$ and $A(\zeta)$ using
(see Ref.~\cite{YangPersson})
$$u_1(\zeta)=\bar u(\zeta)+\bar u'(\zeta) A(\zeta)/A'(\zeta).$$
The quantity $\bar u(\zeta)$ is the average separation between the surfaces in the apparent contact regions
observed at the magnification $\zeta$, see Fig.~\ref{asperity.mag}. 
It can be calculated from\cite{YangPersson}
$$\bar{u}(\zeta ) = \surd \pi \int_{\zeta q_0}^{q_1} dq \ q^2C(q) 
w(q)
 \int_{p(\zeta)}^\infty dp' 
 \ {1 \over p'} e^{-[w(q,\zeta) p'/E^*]^2},$$
where $p(\zeta)=P_0A_0/A(\zeta)$
and
$$w(q,\zeta)=\left (\pi \int_{\zeta q_0}^q dq' \ q'^3 C(q') \right )^{-1/2}.$$
The function $P(q,p,\zeta)$ is given by
$$P(q,p,\zeta) = {2\over \surd \pi} \int_0^{s(q,\zeta)p} dx \ e^{-x^2},$$
where $s(q,\zeta)=w(q,\zeta)/E^*$.

\begin{figure}
\includegraphics[width=0.4\textwidth,angle=0.0]{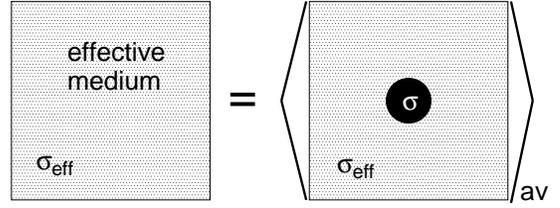}
\caption{\label{effectivemedium}
Effective medium theories take into account random
disorder in a physical system, e.g., in a granular metallic film.
The equation determining the ``effective medium'' (e.g., the effective conductivity $\sigma_{\rm eff}$)
is obtained by calculating
some properties of the effective medium and demanding that the same result
is obtained by embedding in the effective medium a circular region of one component
of the original system, and then averaging over the different component,
with weights determined by the fractional areas of 
the various components in the original physical system.}
\end{figure}

\begin{figure}
\includegraphics[width=0.22\textwidth,angle=0.0]{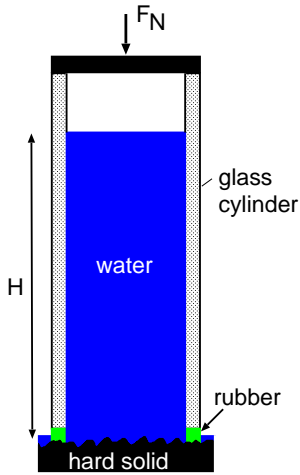}
\caption{\label{tube.water}
Experimental set-up for measuring the leak-rate of seals.
A glass (or PMMA) cylinder with a rubber ring attached to one end is squeezed against
a hard substrate with well-defined surface roughness. The cylinder is filled with 
water, and the leak-rate of the water at the rubber-countersurface
is detected by the change in the height of the water in the cylinder. 
}
\end{figure}


\vskip 0.3cm
{\bf 3. Effective medium theory}

The single-junction theory presented above assumes that the leak-rate is determined by the 
resistance towards fluid flow through the critical constriction. In reality there will be many flow
channels at the interface. Here we will use the 2D Bruggeman effective medium theory
to calculate (approximately) the leak-resistance resulting from the network of flow channels.

Using the 2D Bruggeman effective medium theory we get (see Ref. \cite{Brugg}, 
Fig. \ref{effectivemedium}, and Appendix A):
$$\dot Q = {L_y\over L_x} \sigma_{\rm eff} \Delta P, \eqno(3)$$
where $\Delta P= P_{\rm a}-P_{\rm b}$ is the pressure drop and 
where (see Appendix A)
$${1\over \sigma_{\rm eff}} = \int d\sigma \ P(\sigma) {2\over \sigma_{\rm eff} + \sigma} $$
$$=  
\int d\zeta \left (-{A'(\zeta)\over A_0 }\right )  {2\over \sigma_{\rm eff} + \sigma (\zeta)}, \eqno(4)$$ 
where
$$\sigma (\zeta ) ={[\alpha u_1(\zeta)]^3 \over 12 \eta}. \eqno(5)$$
Eq. (4) is easy to solve by iteration.

It is not clear that the effective medium theory is better than the single-junction theory. One problem
with this theory is the following: In the effective medium model there is no correlation between the size of
a region and the (average) separation between the surfaces in the region. In reality, 
the regions where the surface separation
is large form large compact (or connected) regions
(since they are observed already at low magnification).

\vskip 0.3cm
{\bf 4. Experimental} 

We have performed a very simple 
experiment to test the theory presented above. 
In Fig.~\ref{tube.water} we show our 
set-up for measuring the leak-rate of seals.
A glass (or PMMA) cylinder with a rubber ring (with rectangular cross-section)
attached to one end is squeezed against
a hard substrate with well-defined surface roughness. The cylinder is filled with 
water, and the leak-rate of the fluid at the rubber-countersurface
is detected by the change in the height of the fluid in the cylinder. In this case
the pressure difference $\Delta P = P_{\rm a}-P_{\rm b} = \rho g H$, where $g$ is the gravitation
constant, $\rho$ the fluid density and $H$ the height of the fluid column. With $H\approx 1 \ {\rm m}$
we get typically $\Delta P \approx 0.01 \ {\rm MPa}$. With the diameter of the glass cylinder of
order a few cm, the condition $P_0>> \Delta P$ (which is necessary in order to be able to
neglect the influence on the contact mechanics from the fluid pressure at the rubber-countersurface)
is satisfied already for loads (at the upper
surface of the cylinder) of order kg. In our study we use a   
rubber ring with the Young's elastic modulus $E=2.3 \ {\rm MPa}$, and with the inner and outer diameter
$3 \ {\rm cm}$ and $4 \ {\rm cm}$, respectively, and the height $0.5 \ {\rm cm}$. 
The rubber ring was made from a silicon elastomer (PDMS) prepared using
a two-component kit (Sylgard 184) purchased from Dow Corning (Midland, MI). The kit consist of a base 
(vinyl-terminated polydimethylsiloxane) and a curing agent (methylhydrosiloxane-dimethylsiloxane copolymer) 
with a suitable catalyst. From these two components we prepared a mixture 10:1 (base/cross linker) in weight.
The mixture was degassed to remove the trapped air induced by stirring from the 
mixing process and then poured into casts. The bottom of these casts was made from glass to obtain 
smooth surfaces. The samples were cured in an oven at $80 ^\circ {\rm C}$ for 12 h.

\begin{figure}
\includegraphics[width=0.45\textwidth,angle=0.0]{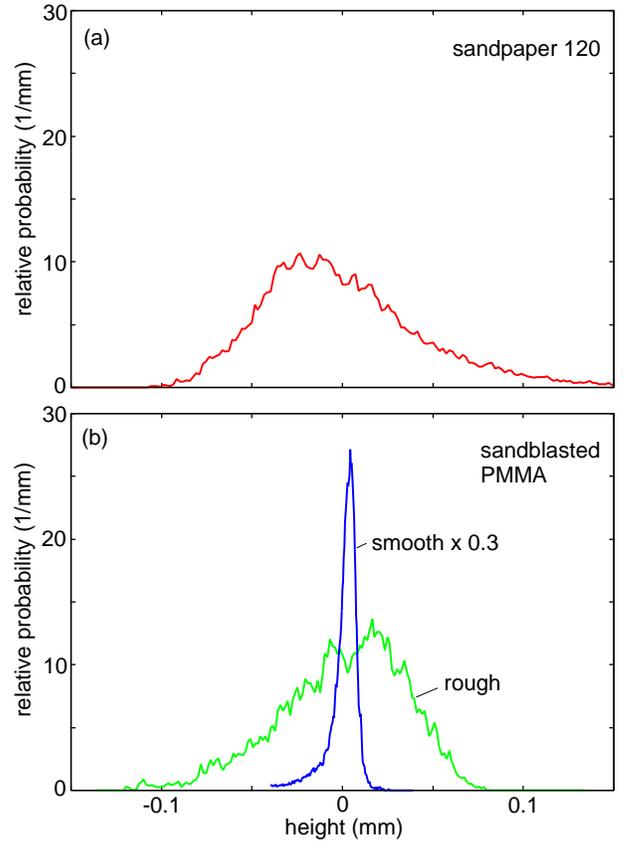}
\caption{\label{all2.plus.sandpaper.Ph}
The surface height probability distribution of (a) sandpaper 120, and (b) two sand-blasted PMMA. The surfaces have the root-mean-square roughness
$44 \ {\rm \mu m}$, $34 \ {\rm \mu m}$, and $10 \ {\rm \mu m}$, respectively, 
and the surface area (including only the surface roughness with wavevector indicated in the figure) 
is about $49 \%$, $28 \%$ and $10\%$ larger, respectively, than the nominal 
surface area $A_0$ (i.e., the surface area projected on the $xy$-plane).}
\end{figure}

\begin{figure}
\includegraphics[width=0.45\textwidth,angle=0.0]{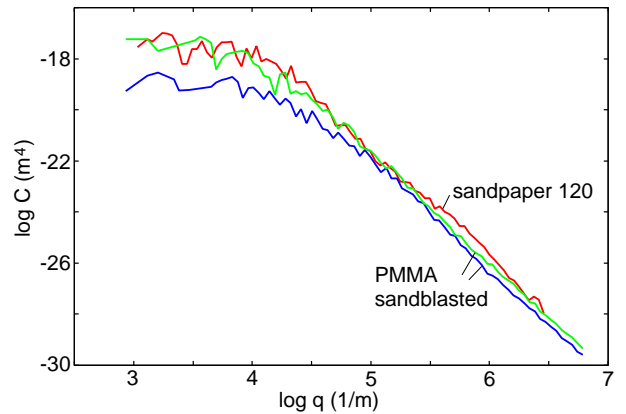}
\caption{\label{Cq.both}
Surface roughness power spectrum of sandpaper 120 and two sand-blasted PMMA. The surfaces have the root-mean-square roughness
$44 \ {\rm \mu m}$, $34 \ {\rm \mu m}$, and $10 \ {\rm \mu m}$, respectively, 
and the surface area (including only the surface roughness with wavevector indicated in the figure) 
is about $49 \%$, $28 \%$ and $10\%$ larger, respectively, than the nominal 
surface area $A_0$ (i.e., the surface area projected on the $xy$-plane).}
\end{figure}

We have used sandpaper and sand-blasted PMMA as substrates. 
The sandpaper (corundum paper, grit size 120) has the root-mean-square roughness $44 \ {\rm \mu m}$.
From the measured surface topography we obtain the height probability distribution $P(h)$ 
and the surface roughness power spectrum $C(q)$
shown in Fig. \ref{all2.plus.sandpaper.Ph}(a) and \ref{Cq.both}, respectively.  
Sand paper has much sharper and larger roughness than the counter surfaces used in normal rubber seal applications.
However, from a theory point of view it should not really matter on which length scale the
roughness occurs, except for ``complications'' such as the influence of adhesion and fluid
contamination particles (which tend to clog the flow channels). 
Nevertheless, the theory assumes that the average surface slope is not too large, and
we have therefore also measured the leak rate for rubber seal in contact with sand-blasted Plexiglas
with less sharp roughness.

Our first experiment with a relative smooth Plexiglas (PMMA) surface
showed that the leak rate decreased by time and finally no leaking could be observed.
But this experiment used unfiltered tap water containing contamination particles which clogged the channels.
Using distilled water we found that the leak rate (for a given fluid pressure difference) to be practically time independent. 
In Fig. \ref{all2.plus.sandpaper.Ph}(b) and \ref{Cq.both}   
we show the height probability distribution $P(h)$ and the
power spectrum $C(q)$ of the two sand-blasted PMMA used below. The root-mean-square roughness of the
two surfaces is $34 \ {\rm \mu m}$ and $10 \ {\rm \mu m}$.
 



\vskip 0.3cm
{\bf 5. Experimental results and analysis}

\begin{figure}
\includegraphics[width=0.45\textwidth,angle=0.0]{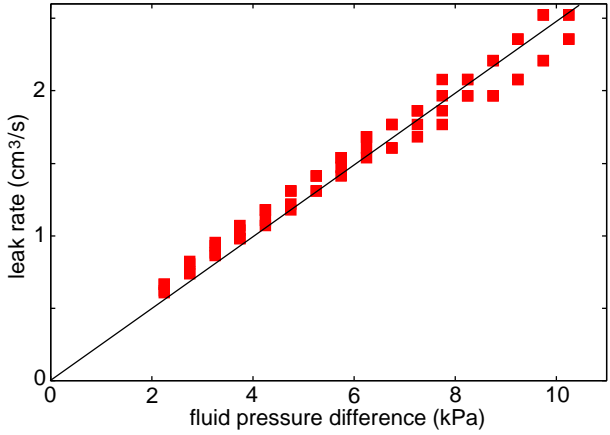}
\caption{\label{pressurelealrate}
Square symbols: the measured leak rate for different fluid pressure drop $\Delta P = P_{\rm a}-P_{\rm b}$ 
for the nominal squeezing pressure $P_0 \approx 60 \ {\rm kPa}$. For the sandpaper surface. 
}
\end{figure}

According to (1) and (3) we expect the leak-rate to increase linearly with the fluid pressure difference
$\Delta P = P_{\rm a}-P_{\rm b}$. We first performed some experiments to test this prediction. In Fig. \ref{pressurelealrate}
we show the measured leak rate (for the sandpaper surface) for different fluid pressure drop $\Delta P$ 
for the nominal squeezing pressure $P_0 \approx 60 \ {\rm kPa}$. To within the accuracy of the experiment, the leak-rate depends linearly on
$\Delta P$.  

\begin{figure}
\includegraphics[width=0.45\textwidth,angle=0.0]{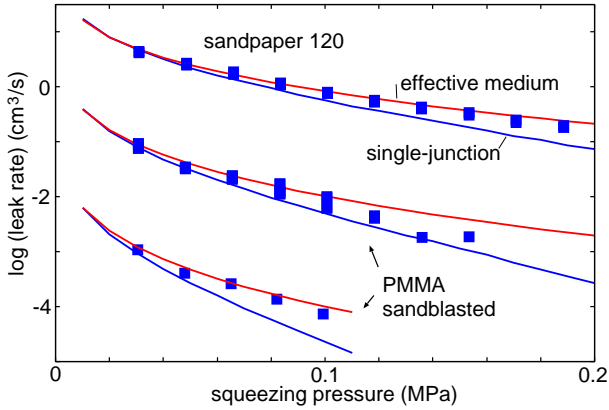}
\caption{\label{Fig.NEW.sandpaperLinescan.factor.0.35.and.0.03.all3.dotQ.pressure}
Square symbols: the measured leak rate for ten different squeezing pressures for 
sandpaper 120 (upper data points) and sand-blasted PMMA (lower two sets of data points). 
In each case the upper solid lines are
the calculated leak rate using the effective medium theory and the lower 
solid lines using the single-junction theory. In the
calculation we used the measured surface topography, the measured rubber elastic
modulus $E=2.3 \ {\rm MPa}$ and the fluid pressure difference 
$\Delta P = P_{\rm a}-P_{\rm b} = 10 \ {\rm kPa}$ obtained from the height of the water column.  
In the calculations we have used $\alpha = 0.7$ (for sandpaper) and $0.3$ (for PMMA).
}
\end{figure}

In Fig. \ref{Fig.NEW.sandpaperLinescan.factor.0.35.and.0.03.all3.dotQ.pressure} 
we show the logarithmic (with 10 as basis) of the measured leak rate for several different squeezing pressures
(square symbols). We show results for both the sandpaper surface and for the two sand-blasted PMMA surfaces.
The solid lines are the calculated leak rate using the measured rubber elastic modulus $E=2.3 \ {\rm MPa}$ and the 
surface power spectrum $C(q)$ shown in Fig. \ref{Cq.both}.
We show calculations using both the single-junction theory (blue lines) and the effective medium theory (red lines).
In the calculations we have used $\alpha = 0.7$ (for sandpaper) and $0.3$ (for PMMA).
Note that both theories gives similar results for low squeezing pressures, but for larger squeezing pressures the effective medium theory
gives a larger leak rate. The experimental data agree better with the effective medium theory than with the single-junction theory.



\begin{figure}
\includegraphics[width=0.35\textwidth,angle=0.0]{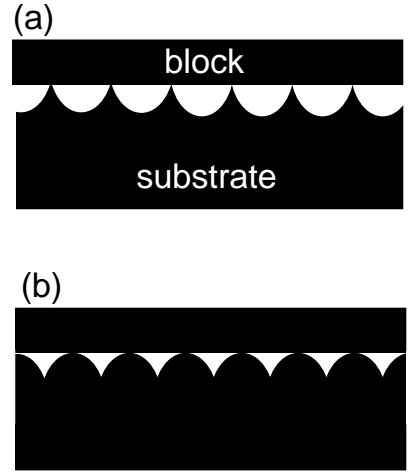}
\caption{\label{twotypes}
Contact between a rigid block with a flat surface and a rigid substrate
with periodic surface structures.
Two substrate surfaces in (a) and (b) have the same surface roughness power spectrum.
Note that the empty volume between the surfaces is much larger in the case (a) than
in case (b).}
\end{figure}

\vskip 0.3cm
{\bf 6. Discussion}

We have presented experimental results for the leak rate for a PDMS rubber
ring (with rectangular cross section), squeezed against three different 
surfaces: two sand-blasted PMMA surfaces and a sandpaper 120 surface. The experimental results
have been compared to a simple single-junction theory and to a (more accurate) effective medium
theory. The basic input in both theories is information about the interfacial surface separation,
which we have obtained using the contact mechanics theory of Persson. The pressure dependence
predicted by the theory is in good agreement with the experimental data, in particular for the
effective medium theory. 

The contact mechanics theory we use assumes randomly rough surfaces. Randomly rough surfaces have
a Gaussian height probability distribution $P(h)$. However, most surfaces of engineering interest
have not Gaussian height probability 
distribution. In Fig. \ref{all2.plus.sandpaper.Ph} we show the 
height distribution for the surfaces used in the present study. Note that $P(h)$ is asymmetric 
with a tail towards higher $h$ for the sandpaper surface, and towards smaller (negative) $h$ for
the sand-blasted PMMA surfaces. This is easy to understand: the sandpaper surfaces consist of
particles with sharp edges pointing above the surface, while the region between 
particles are filled with a resin-binder making the valleys smoother and wider than the peaks, which result
in an asymmetric $P(h)$ as observed. The PMMA surfaces are prepared by bombarding a flat
PMMA surface with small hard particles. This result, at least for short time of sand-blasting, 
in local indentations (where the particles hit the surface) separated by smoother surface regions,
leading to the observed asymmetry in the height distribution.

Let us now discuss how the asymmetry in the height distribution may effect the leak rate. 
To illustrate this we consider an extreme case: a rigid solid block with a flat surface in contact
with a rigid substrate with periodic ``roughness'' as in Fig. \ref{twotypes}. 
The substrate surfaces in (a) and (b) have the same surface roughness power spectrum, but it
is clear that in (a) the empty volume between the surfaces is larger than in (b), resulting 
in a larger leak rate. In the real situation the roughness is not periodic and the solids are
not rigid, but one may expect a higher leak rate for the situation
where the asymmetry of the height profile is as for the sandpaper surface. We suggest that
this may be the physical origin of why the factor $\alpha$ is larger for the sandpaper surface
as compared to the PMMA surfaces. Another observation which support this conclusion
is the fact that the surface roughness power spectrum of the rough PMMA surface and the 
sandpaper 120 surface are very similar, but the leak rate differ by roughly two orders
of magnitude. This indicate that some aspects of the surface topography, not contained in the power spectrum, 
is likely to be important. We note that for randomly rough surface, the statistical properties of the surfaces are
fully contained in the power spectrum $C(q)$, i.e., for this case only $C(q)$ will enter in the 
theory for the leak-rate.

To study the point discussed above, we plan to perform an experiment where 
we ``invert'' the roughness of the sandpaper surface by producing
a ``negative'' using silicon rubber. In this experiment we will squeeze a silicon rubber ring, which is 
cross-linked with the sandpaper surface as the substrate, against a flat glass surface. 
By comparing the measured leak-rate for this configuration with that for a silicon ring with
flat bottom surface squeezed against the same sandpaper surface, we will be able to address the
problem illustrated in Fig. \ref{twotypes}.

An alternative to using the effective medium approach to calculate the leak-rate
of seals, one may use the so called critical path analysis\cite{Langer}. This approach has recently been applied to
seals\cite{Bott} but in contrast to the effective medium theory, there enters two 
parameters which are not easy to obtain from theory. Since the effective medium approach
has been found to be rather accurate (see, e.g., Ref. \cite{Kirk}) we believe this approach is more suitable for calculating
the leak rate of seals.

\vskip 0.3cm
{\bf 7. Summary and conclusion}

To summarize, we have compared 
experimental data with theory for the leak-rate of seals. The theory is 
based on percolation theory and a recently developed contact mechanics theory.
The experiments are for silicon rubber seals in contact with sandpaper and two sand-blasted PMMA surfaces. The
elastic properties of the rubber and the surface topography of the 
sandpaper and PMMA surfaces are fully characterized.
The dependence of the calculated leak-rate $\dot Q$ on the squeezing pressure 
is in good agreement with experiment.
The simplest version of the
theory only account for fluid flow through the percolation channels
observed at (or close to) the percolation threshold. We have also presented
another approach based on the effective medium approximation. This theory 
also include flow channels observed at higher magnification, and gives
larger leak-rates than the single-junction theory, 
which only includes one leak-rate channel
(or $L_y/L_x$ channels for a rectangular seal). 

\vskip 0.3cm

{\bf Acknowledgments}

We thank G. Carbone for useful comments on the manuscript.
This work, as part of the European Science Foundation EUROCORES Program FANAS, was supported from funds 
by the DFG and the EC Sixth Framework Program, under contract N ERAS-CT-2003-980409.

\vskip 0.5cm

{\bf Appendix A}

Here we briefly review the effective medium approach for calculating the
fluid flow through an interface where the separation $u({\bf x})$ between the surfaces varies with
the lateral coordinate ${\bf x} = (x,y)$. 
If $u({\bf x})$ varies slowly with ${\bf x}$ the Navier-Stokes equations of fluid flow reduces to
$${\bf J} = -\sigma \nabla p \eqno(A1)$$
where the conductivity $\sigma = u^3({\bf x})/12 \eta$. 

In the effective medium approach one replace the local, spatial varying, conductivity $\sigma ({\bf x})$
with a constant effective conductivity $\sigma_{\rm eff}$. Thus the fluid flow current equation
$${\bf J} = -\sigma_{\rm eff} \nabla p, \eqno(A2)$$
as applied to a rectangular region $L_x\times L_y$ with the 2D pressure gradient $dp/dx = (P_{\rm b} - P_{\rm a})/L_x$,
gives
$$\dot Q = L_y J_x = {L_y\over L_x} \sigma_{\rm eff} \Delta P \eqno(A3)$$
where $\Delta P = P_{\rm a} - P_{\rm b}$ is the pressure drop.

The effective medium conductivity $\sigma_{\rm eff}$ is obtained as follows.
Let us study the current flow at a circular inclusion (radius $R$) with 
the (constant) conductivity $\sigma$ located
in an infinite conducting sheet with the (constant) conductivity $\sigma_{\rm eff}$. 
We introduce polar coordinates with the origin at the center of
the circular inclusion. The current
$${\bf J} = -\sigma \nabla p \ \ \ \ \ {\rm for} \ \ \ \ \ r<R$$
$${\bf J} = -\sigma_{\rm eff} \nabla p \ \ \ \ \ {\rm for} \ \ \ \ \ r > R$$
We consider a steady state so that
$$\nabla \cdot {\bf J} = 0$$
or
$$\nabla^2 p = 0\eqno(A4)$$
If ${\bf J}_0$ is the current far from the inclusion (assumed to be constant) we get
for $r > R$:
$$p = \left [1+f(r)\right ] {\bf J}_0\cdot {\bf x}\eqno(A5)$$ 
Eq. (A4) is satisfied if
$$f''(r) + 3 f'(r) r^{-1} = 0$$
A solution to this equation is $f=\alpha r^{-2}$. Substituting this in (A5) gives
$$p=\left [1+\alpha r^{-2} \right ] {\bf J}_0\cdot {\bf x} \eqno(A6)$$
For $r < R$ we have the solution
$$p=\beta {\bf J}_0\cdot {\bf x} \eqno(A7)$$
Since $p$ and ${\bf x}\cdot {\bf J}$ must be continuous at $r=R$ we get from (A6) and (A7):
$$1+\alpha R^{-2} = \beta $$
$$\left (1-\alpha R^{-2} \right ) \sigma_{\rm eff} = \beta \sigma$$
Combining these two equations gives
$$\beta ={2\sigma_{\rm eff} \over \sigma_{\rm eff}+\sigma } \eqno(A8)$$
The basic picture behind effective medium theories is presented in Fig. \ref{effectivemedium}.
Thus, for a two component system, 
one assumes that the flow in the effective
medium should be the same as the average fluid flow obtained 
when circular regions of the two components are embedded in the effective medium.
Thus, for example, the pressure $p$ calculated assuming that the effective medium occur everywhere
must equal the average $c_1p_1+c_2p_2$ of the pressures $p_1$ and $p_2$ calculated with the 
circular inclusion of the two components {\bf 1} and {\bf 2}, respectively. For $r < R$
we have for the effective medium $p={\bf J}_0\cdot {\bf x}$ and using (A7) 
the equation $p=c_1p_1+c_2p_2$ gives
$$1= c_1 \beta_1 + c_2 \beta_2\eqno(A9)$$
where $c_1$ and $c_2$ are the fractions of the total area occupied by
the components {\bf 1} and {\bf 2}, respectively. Using (A8) and (A9) gives
$$1= c_1 {2\sigma_{\rm eff} \over \sigma_{\rm eff}+\sigma_1 } + c_2 {2\sigma_{\rm eff} \over \sigma_{\rm eff}+\sigma_2 }$$
which is the standard Bruggeman effective medium for a two component system.

If one instead have a continuous distribution of components (which we number by the continuous index $\xi$) 
with conductivities $\sigma = \sigma (\xi)$, then 
$$1= \int d\xi \ P(\xi) \beta(\xi) \eqno(A10)$$
where $P(\xi)$ is the fraction of the total surface area occupied by the 
component denoted by $\xi$. The probability distribution $P(\xi)$ is
normalized so that 
$$\int d\xi \ P(\xi)=1 \eqno(A11)$$
Using (A8) we get
$$1= \int d\xi \ P(\xi) {2\sigma_{\rm eff} \over \sigma_{\rm eff}+\sigma (\xi) } \eqno(A12)$$


\vskip 0.5cm

\begin{thebibliography}{99}

\bibitem{Flitney}
R. Flitney, {\it Seals and sealing handbook} (Elsevier, 2007). 

\bibitem{Mofidi}
M. Mofidi, B. Prakash, B.N.J. Persson and O. Albohr,
J. Phys.: Condens. Matter {\bf 20}, 085223 (2008).

\bibitem{P3}
See, e.g., B.N.J. Persson, O. Albohr, U. Tartaglino, A.I. Volokitin and E. Tosatti, 
J. Phys. Condens. Matter {\bf 17}, R1 (2005). 

\bibitem{LorenzEPL}
B. Lorenz and B.N.J. Persson, EPL {\bf 86}, 44006 (2009).

\bibitem{Creton}
B.N.J. Persson, O. Albohr, C. Creton and V. Peveri, 
J. Chem. Phys. {\bf 120}, 8779 (2004)

\bibitem{Yang}
B.N.J. Persson and C. Yang, J. Phys.: Condens. Matter, {\bf 20}, 315011 (2008)

\bibitem{JCPpers}
B.N.J. Persson, J. Chem. Phys. {\bf 115}, 3840 (2001).

\bibitem{PerssonPRL}
B.N.J. Persson, Phys. Rev. Lett. {\bf 99}, 125502 (2007).

\bibitem{PSSR}
B.N.J. Persson, Surf. Science Reports {\bf 61}, 201 (2006).

\bibitem{P1}
B.N.J. Persson, Eur. Phys. J. E{\bf 8}, 385 (2002).

\bibitem{Bucher} 
B.N.J. Persson, F. Bucher and B. Chiaia, Phys. Rev. B{\bf 65}, 184106 (2002).

\bibitem{YangPersson}
C. Yang and B.N.J. Persson, J. Phys.: Condens. Matter {\bf 20}, 215214 (2008).

\bibitem{PerssonJPCM}
B.N.J. Persson, J. Phys.: Condens. Matter {\bf 20}, 312001 (2008). 

\bibitem{earlier} 
The contact mechanics model developed in 
Ref. \cite{JCPpers,PerssonPRL,PSSR,P1,Bucher,YangPersson,PerssonJPCM}
takes into account the elastic 
coupling between the contact regions in the nominal rubber-substrate contact area.
Asperity contact models, such as the ``standard'' 
contact mechanics model of Greenwood--Williamson\cite{GW}, and the 
model of Bush et al\cite{Bush},
neglect this elastic coupling, which results in highly 
incorrect results\cite{Carlos,Carbone}, 
in particular for the relations between the squeezing pressure and the
interfacial separation\cite{Lorenz}.
 
\bibitem{GW}
J.A. Greenwood and J.B.P. Williamson, Proc. Roy. Soc. London A{\bf 295}, 300 (1966).

\bibitem{Bush}
A.W. Bush, R.D. Gibson and T.R. Thomas, Wear {\bf 35}, 87 (1975).

\bibitem{Carlos}
C. Campana, M.H. M\"user and M.O. Robbins, J. Phys.: Condens. Matter {bf 20}, 354013 (2008)

\bibitem{Carbone}
G. Carbone and F. Bottiglione, J. Mech. Phys. Solids {\bf 56}, 2555 (2008).

\bibitem{Lorenz}
B. Lorenz and B.N.J. Persson, J. Phys.: Condens. Matter {\bf 201}, 015003 (2009).

\bibitem{Stauffer}
D. Stauffer and A. Aharony, {\it An Introduction to Percolation Theory}, CRC Press (1991).

\bibitem{complex} 
Fig. \ref{pic.Az.Azdz.rough}(a) is schematic as in reality
the contact islands at high enough magnification are fractal-like, and decreasing the magnification
result in more complex changes than just adding strips (of constant width) of contact area to 
the periphery of the contact islands. However, this does not change our conclusions.

\bibitem{WithYang}
In Ref. \cite{YangPersson} the probability distribution of 
interfacial separations $\langle \delta (u-u({\bf x})) \rangle$
as obtained from Molecular Dynamics calculations for self-affine fractal surfaces
(with the fractal dimension $D_{\rm f}=2.2$) was compared to the distribution of separations obtained from
$u_1(\zeta)$. The former distribution was found to be about a factor of two wider than that obtained from $u_1(\zeta)$.
This is consistent with the fact that $u_1(\zeta)$ is already an averaged separation and indicate that in this case
$\alpha \approx 0.5$.

\bibitem{Brugg}
D. Bruggeman, Ann. Phys. Leipzig {\bf 24}, 636 (1935).

\bibitem{Langer}
V.N. Ambegaokar, B.I. Halperin and J.S. Langer, Phys. Rev. B{\bf 4}, 2612 (1971);
A.G. Hunt, {\it Percolation Theory for Flow in Porous Media} (Springer, New York, 2005);
Z. Wu, E. Lopez, S.V. Buldyrev, L.A. Braunstein, S. Havlin and H.E. Stanley, 
Phys. Rev. E{\bf 71}, 045101(R) (2005).

\bibitem{Bott}
F. Bottiglione, G. Carbone, L. Mangialardi and G. Mantriota, J. Applied Physics {\bf 106}, xxx (2009).

\bibitem{Kirk} 
S. Kirkpatrick, Reviews of Modern Physics {\bf 45}, 574 (1973).

\end{thebibliography}
\end{document}